\documentclass[%
 aps,
 sd,%
 amsmath,amssymb,
 reprint,%
]{revtex4-1}

\usepackage{graphicx}  
\usepackage{dcolumn}   
\usepackage{bm}        
\usepackage{graphicx}
\usepackage{subfig}
\usepackage{epsfig}
\usepackage{epstopdf}
\usepackage{physics}
\usepackage{enumerate}
\newcommand{\RNum}[1]{\uppercase\expandafter{\romannumeral #1\relax}}

\usepackage{url}

\begin{document}

\preprint{AIP/123-QED}

\title{Role of Initial Coherence in Excitation Energy Transfer in Fenna-Matthews-Olson Complex}

\author{Davinder Singh}
\email{davinder.singh@iitrpr.ac.in}

\author{Shubhrangshu Dasgupta}%

\affiliation{Department of Physics, Indian Institute of Technology Ropar, Rupnagar, Punjab - 140001, India}

\date{\today}   


\begin{abstract}
We theoretically show that the initial coherence plays a crucial role in enhancing the speed of excitation energy transfer (EET) in Fenna-Matthews-Olson (FMO) complex. We choose a simplistic eight-level model considering all the bacateriochlorophyll-a sites in a monomer of FMO complex.  We make a comparative numerical study of the EET, in terms of non-Markovian evolution of an initial coherent superposition state and a mixed state. A femto-second coherent laser pulse is suitably chosen to create the initial coherent superposition state. Such an initial state relaxes much faster than a mixed state thereby speeding up the EET. In this analysis, we have taken into account the relative orientation of the transition dipole moments of the bacateriochlorophyll-a sites and their relative excitation energies. Our results reveal that for 2D electronic spectroscopy experiments, the existing two-pathway model of energy transfer in FMO complex may not be suitable in our understanding of EET.
\end{abstract}

\maketitle


\section{Introduction}

In bacterial photosynthesis, antenna complexes harvest the photon and transfer the captured energy to the reaction center (which stores it chemically) with almost near unity quantum efficiency \cite{chain_quantum_1977,Amerongen,Blankenship}. Resolving the functioning of these complex biological systems could be useful for newer solar energy applications \cite{Meyer_Chemist_2011}.

During the bacterial photosynthesis in, e.g., {\it Chlorobium Tepidum}, a photon is captured {\it in situ} by a light-harvesting molecule of the chlorosome antenna and this excitation is then transferred to the Fenna-Matthews-Olson (FMO) complex. This complex is a trimer and works as a energy transmitting wire between the chlorosome antenna and the reaction center \cite{Amerongen,Fenna_Chlorophyll_1975,milder_revisiting_2010}. Each monomer in this trimer consists of seven bacteriochlorophyll-a (BChla) molecules immersed in protein environment and transmits the excitation energy quite independently from the other monomers due to weak Coulomb coupling among them \cite{Adolphs,Ishizaki}. The BChla 1 and BChla 6 of each monomer are close to the base-plate of the chlorosome antenna and the BChla 3 and BChla 4 are close to reaction center\cite{Wen}. 

Recent experimental observations, based on 2D electronic spectroscopy (2-DES) \cite{Brixner,Brixner2,Cho} on the {\it isolated} FMO complex, have revealed that the electronic excitation gets transferred in a wave-like manner \cite{Engel_evidence_2007,Panitchayangkoon_long_2010,
panitchayangkoon_direct_2011} rather than incoherent hopping, between different BChla sites. These long lasting ($\sim 600 fs $) oscillations were arguably explained as reminiscence of coherent beats (or dynamical coherence). These coherence's were further explained as originated from electronic modes and it was argued that coherence avoids the energetic traps and enhances the efficiency of EET\cite{Ishizaki,Nalbach,palmieri_lindblad_2009,bhattacharyya_adiabatic_2013,ai_complex_2012,
shabani_efficient_2012,chen_rerouting_2013,hoyer_spatial_2012}. The inhomogeniety of protein environment and its effect on coherence were also illustrated to study the EET dynamics\cite{rivera,mohseni_environment-assisted_2008,caruso_highly_2009,eisfeld_classical_2012,mujica-martinez_quantification_2013,sarovar_environmental_2011}. Recently the origin of dynamical oscillatory coherence was explained on the basis of vibronic modes of BChla sites \cite{chenu_enhancement_2013,kreisbeck_disentangling_2013,
 tempelaar_vibrational_2014,christensson_origin_2012,chin_role_2013,killoran_enhancing_2015}. However so far the effect of initial coherence created by the coherent laser light as used in 2-DES experiments has not been explored.

In the 2-DES experiments, the FMO complex is extracted {\it out} from the bacteria (namely, {\it Chlorobium Tepidum}). An ultra-short femtosecond laser pulse is used to excite it; thereby it creates a coherent superposition of several excited BChla sites. However in a realistic bacterial EET,  the incoherent excitation (due to sunlight) makes an incoherent mixture of the excitations of the BChla sites. Different initial conditions are expected to have a substantially varied effect on the temporal dynamics of EET \cite{cao_correlations_2006}. In the most of the earlier theoretical models \cite{Ishizaki,prior_efficient_2010,chin_role_2013} to explain the observations of 2-DES experiments, either the BChla 1 or the BChla 6 has been considered as initially excited. The dynamics thereby follows two different pathways, initiated by excitation in BChla 1 and BChla 6 respectively to reach the destination BChla's (namely BChla 3 and BChla 4). However, such initial conditions clearly do not match with that of the laser excited FMO complex or {\it in-vivo} absorption of excitation by BChla sites of monomer. In this paper, we suitably choose the initial condition and explicitly study the effect of initial coherence on the EET, specifically by comparing the EET for initial coherent and incoherent excitation. Through the study of the dynamics of coherently initiated EET one would be able to mimic the observations of the 2-DES experiments, which are still not fully understood.  We first describe how a superposition of the excited BChla sites is created when a short pulse is applied. We compare its dynamics with that of mixed state that is formed by the incoherent absorption of light. 

Our numerical study has the following salient features:

\begin{enumerate}[(i)]
\item We propose that the distribution of initial excitation is not dictated by the local minima of excited states energy rather by the relative orientation of the dipole moments and the polarization of laser pulse. This would govern the preferential distribution of excitation among all the BChla sites.

\item X-ray crystallography of the FMO complex reveals an inhomogeneous protein environment of different BChla sites. Further, various vibronic modes of BChla sites also effect the EET dynamics. We have incorporated these features by fitting the spectral density with experimentally observed phonon wing and vibronic peaks\cite{singh_coherence_2017}. Moreover, following Wendling {\it et al}\cite{wendling_electronvibrational_2000}, we have used five different vibronic modes with large value of Franck-Condon factors instead of using only one or two modes as has been done in earlier approaches \cite{chenu_enhancement_2013,kreisbeck_disentangling_2013,
 tempelaar_vibrational_2014,christensson_origin_2012,chin_role_2013}.

\item We have carefully considered all the realistic parameters for each BChla site, e.g., their excited state energy and mutual Coulomb couplings.
\end{enumerate}
 
 The structure of the paper is as follows: In section \RNum{2} we describe the preparation of initial state. In the following section (Section \RNum{3}), we illustrate the EET dynamics for initial coherent superposition as well as initial incoherent mixed state in FMO complex. Finally in section \RNum{4} we conclude with summary and an outlook.


\section{Preparation of initial state}

\begin{figure}[!h]
\begin{center}
\subfloat{\label{FMO_Angle}
\includegraphics[width = 2.1 in]{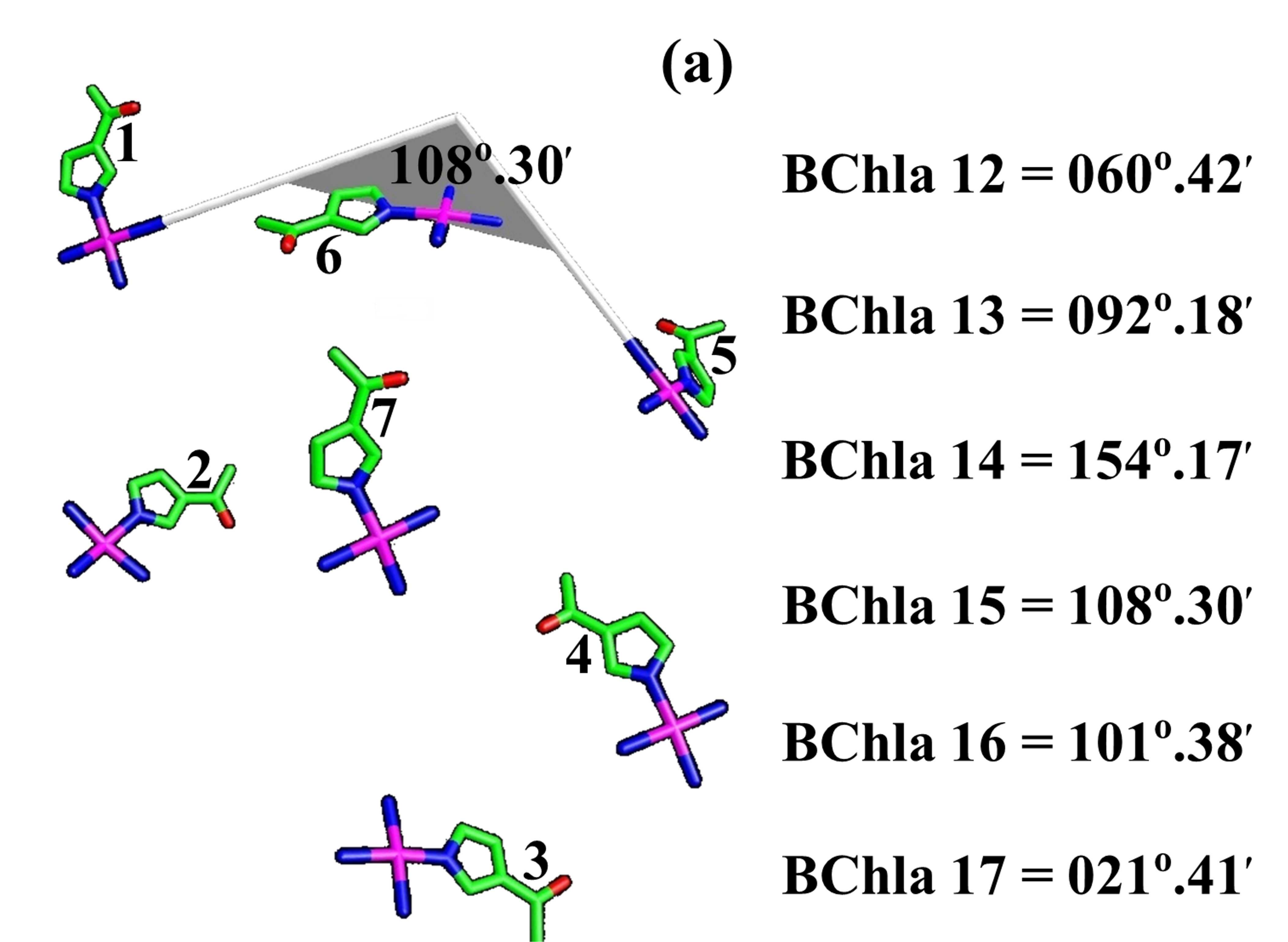}}
\quad
\subfloat{\label{Seven_Levels}
\includegraphics[width = 3 in]{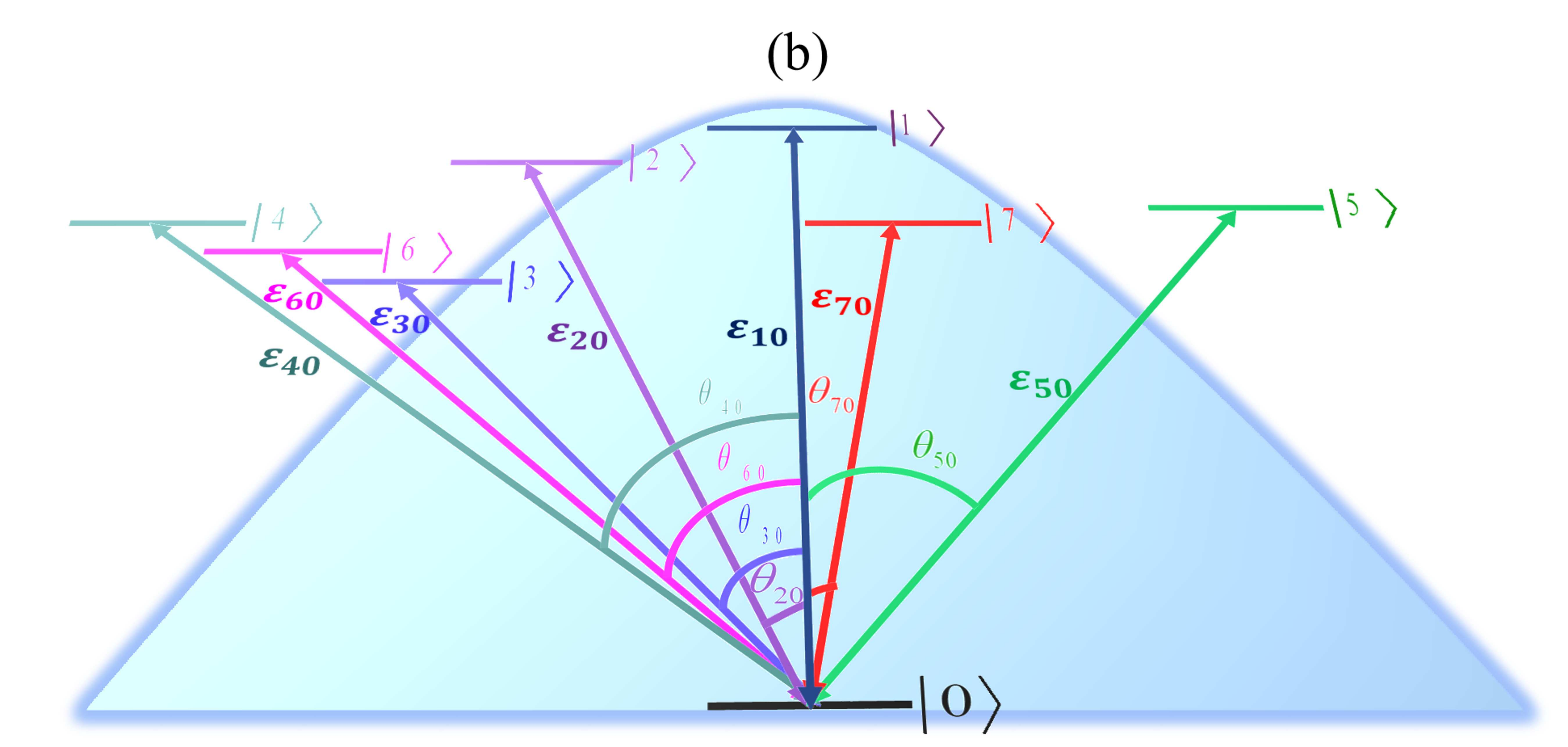}}
\caption{(Color online)(a) Orientation of transition dipole moments of different BChla sites. This figure is created with VEGA-ZZ \cite{pedretti_vega_2004} using PDB entry 3EOJ. (b) Schematic illustration of interaction of Gaussian laser pulse with different BChla sites in a monomer of the FMO complex. Here $\epsilon_{j0}$ represents the transition frequency of the $j$th BChla site and $\theta_{j0}$ is the angle between polarization of light pulse and the transition dipole moment of the $j$th BChla site.}
\label{Angle_Seven_Levels}
\end{center}
\end{figure}

To understand the coherent distribution of excitation among the different BChla sites, each BChla site is modelled as a two-level system. We assume that the EET happens only in the single-excited BChla basis  (see Fig. \ref{Seven_Levels}), i.e., the probabilities of bi-exciton and other higher states are negligible compared to the single excitation.  Such suppression of bi-exciton pathways is also preferable to avoid damage due to excess energy \cite{dong_photon_2017} and can be attributed to dipole blockade arising from the closely-spaced arrangement of the BChla sites\cite{weidemuller_rydberg_2009}. Also note that, in the 2-DES experiments,  the temporal dynamics of only the excitons 
 have been observed \cite{Brixner,Brixner2,Cho}. 

To excite the BChla sites in the monomer of the FMO complex, a linearly polarized pulse $\vec{E}(t) = \hat{\epsilon}\varepsilon(t) e^{-i\omega_{L}t} + {\rm h.c.}$ of the central frequency $\omega_{L}$ and the time-dependent amplitude $\varepsilon(t)$ is applied. The Hamiltonian in the dipole-approximation can therefore be written as

\begin{equation}
\begin{array}{lll}
H &=& \sum_{j = 1}^{7}\left\{\hbar \epsilon_{0}\ket{0}\bra{0} + \hbar \epsilon_{j}\ket{j}\bra{j}\right.\\
&+&\sum_{i>j} \Delta_{ij}\left( \ket{i}\bra{j} + \ket{j}\bra{i}\right) \\
&-&\left.\left(\vec{d}_{j0}\ket{j}\bra{0} + \vec{d}_{0j}\ket{0}\bra{j}\right).\vec{E}\right\} \;.
\end{array}
\label{Hamiltonian}
\end{equation}
Here $\epsilon_{j}$ represents the unperturbed energy of the $j$th BChla site and $\Delta_{ij}$ is the tunneling frequency between $i$th and $j$th BChla site. The transition dipole moment matrix element of the transition $\ket{j}\leftrightarrow\ket{0}$ is represented by $\vec{d}_{j0}$. Here we assume that the time-scale of the laser pulse is much shorter than the time-scale at which the coupling of the bath modes become effective.  The above pulse would create a general superposition of all the relevant states of the monomer. Considering a common ground state $|0\rangle$ and seven excited states $|j\ne 0\rangle$, such a superposition can be written as $\ket{\psi} = \sum_{j = 0}^{7}c_{j}(t)\ket{j}$. 
In the interaction picture, this would evolve with time according to the following Schrodinger equation:
\begin{equation}
\frac{d\ket{\psi}}{dt} = -\dfrac{i}{\hbar}H_{I}\ket{\psi} \;.
\label{Master Eq}
\end{equation}
where $H_{I}$ represents the interaction Hamiltonian. The probability amplitude $c_j$ for the excited state of the $j$th site therefore evolves according to the following equation:
\begin{equation}
\small
\dot{c}_{j} = -i2G(t)c_{0}\cos\theta_{j0}\cos\omega_{L}te^{i(\epsilon_{j} - \epsilon_{0})t} - \sum_{i}i \Delta_{ij}c_{i}e^{i(\epsilon_{j} - \epsilon_{i})t}  \qquad j\neq0\;.
\label{Population}
\end{equation}

We choose the following Gaussian profile of the Rabi frequency of the pulse:                                                                                                                                                                                                                                                                                                                                                                                                                                                                                                                                                                                                                                                                                                                                                                                                                                                                                                                                                                                                                                                                                                                                                                                                                                                                                                                                                                                                                                                                                                                                                                                                                                                                                                                                                                                                                                                                                                                                                                                                                                                                                                                                                                                                                                                                                                                                                                                                                                                                                                                                                                                                                                                                                                                                                                                                                                                                                                                                                                                                                                                                                                                                                                                                                                                                                                                                                                                                                                                                                                                                                                                                                                                                                                                                                                                                                                                                                                                                                                                                                                                                                                                                                                                                                                                                                                                                                                                                                                                                                                                                                                                                                                                                                                                                                                                                                                                                                                                                                                                                                                                                                                                                                                                                                                                                 $G(t) = ge^{-\dfrac{(t - t_{0})^{2}}{\Delta\tau^{2}}}$. We assume that the transition dipole moments $\vec{d_{j0}}$ of the BChla sites have approximately the same magnitude $d$, such that $g = \dfrac{d\varepsilon}{\hbar}$. However, these dipoles have different orientations in three dimension, designated by the angle $\theta_{j0}$ it makes with the polarisation direction of the pulse. We assume that the polarisation is parallel to the transition dipole moment of BChla 1 as illustrated in Fig. \ref{Seven_Levels}, while the other dipoles are oriented at non-zero angles. To analyse the dynamics of interaction, the angles of transition dipole moments are calculated using VEGA-ZZ as illustrated in Fig. \ref{FMO_Angle} \cite{pedretti_vega_2004}. Further, different values of the  transition frequencies of the different BChla sites, as indicated by Adolphs and Renger \citep{Adolphs}, have been considered in our numerical study.  In the view of average transition frequency, we choose $\omega_{L} = 12420 {\rm cm}^{-1}$.

We consider that the laser pulse interacts with the FMO complex for a femtosecond time scale ($\sim 40 fs$). However, the effect of tunneling rate $\Delta_{ij}$ will be dominant at a time scale $\frac{1}{\Delta_{ij}}$ (ranging from $0.4 ps$ to $3 ps$), much after the femtosecond pulse dies away. To elucidate this effect, we numerically solve the set of Eqs.(\ref{Population}) for the probability amplitude of BChla 1 and BChla 2 only, which have the strongest Coulomb coupling among all the combinations of the BChla sites. We observe that for first 40 fs, when the laser pulse interacts with the BChla's, the dynamics remains unaffected by the Coulomb coupling (i.e $\Delta_{ij}$) as illustrated by lower panel of Fig.\ref{Pulse_Population}. However, after the pulse dies out, the effect of $\Delta_{ij}$ becomes dominant as it creates the oscillations of population between two BChla's. Based on this observation, we neglect the effect of Coulomb coupling $\Delta_{ij}$ for the dynamics of full monomer FMO complex for the interval during which the laser pulse interacts.

\begin{figure}[!h]
\begin{center}
\includegraphics[width = 3 in]{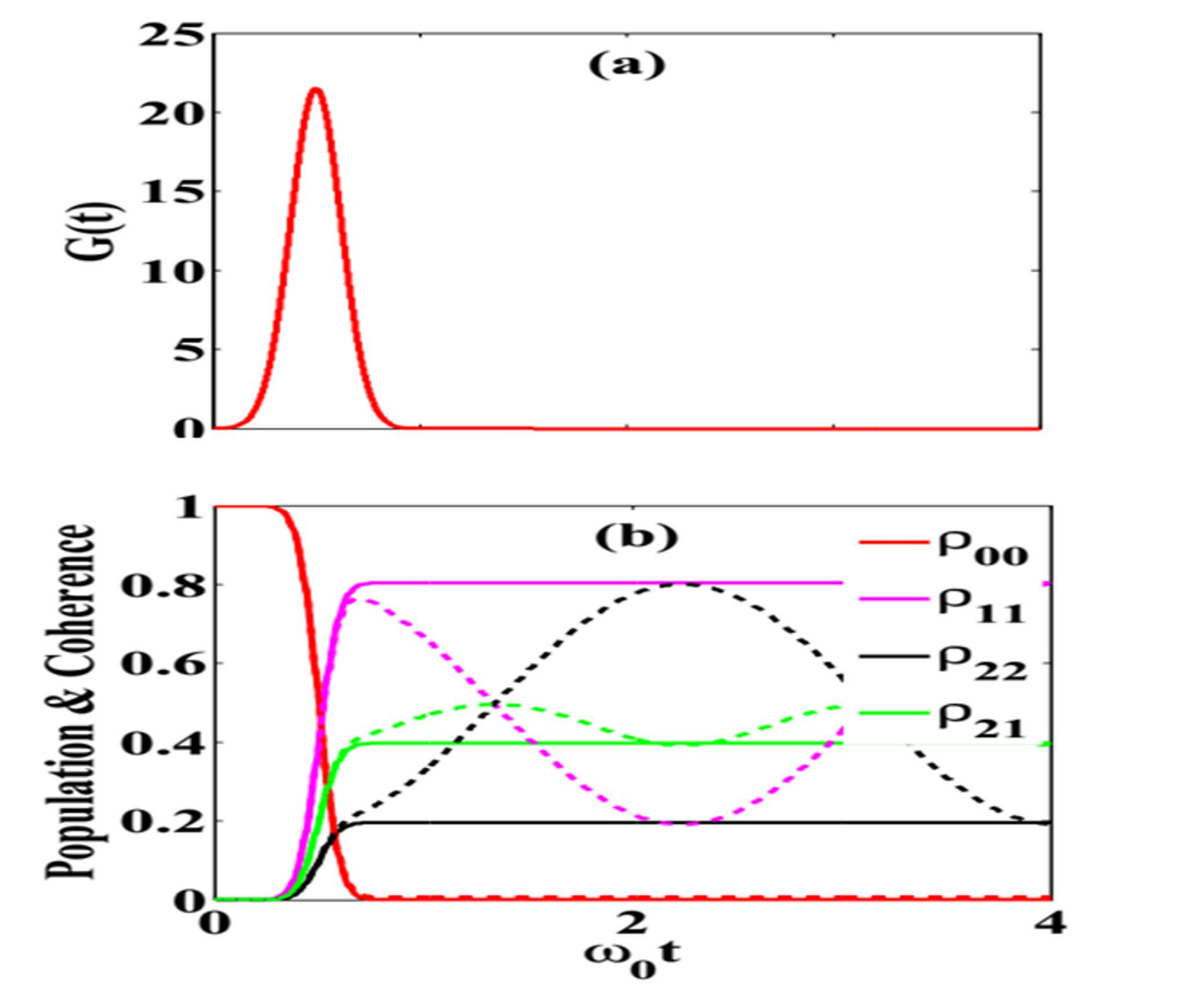}
\caption{(Color online) Temporal dynamics of interaction of laser pulse with two BChla sites (BChla 1 and BChla 2). (a) Profile of a Gaussian pulse of width 15 $fs$. (b) Time evolution of population distribution among two BChla sites. Here the solid line represents the absence of Coulomb coupling (i.e $\Delta_{21} = 0$), while dotted line illustrate the dynamics in the presence of Coulomb coupling (i.e $\Delta_{21} = 87.7 cm^{-1}$). We use normalized parameters $g = 21.5$, $t_{0} = 0.8$ and $\Delta\tau^{2} = 0.03$ and choose $\omega_{0}$ = 100 {\rm cm}$^{-1}$ for normalization.}
\label{Pulse_Population}
\end{center}
\end{figure}

\begin{figure*}[!h]
\begin{center}
\subfloat{\label{Initial_Condition_Coherence}
\includegraphics[width = 2.0 in]{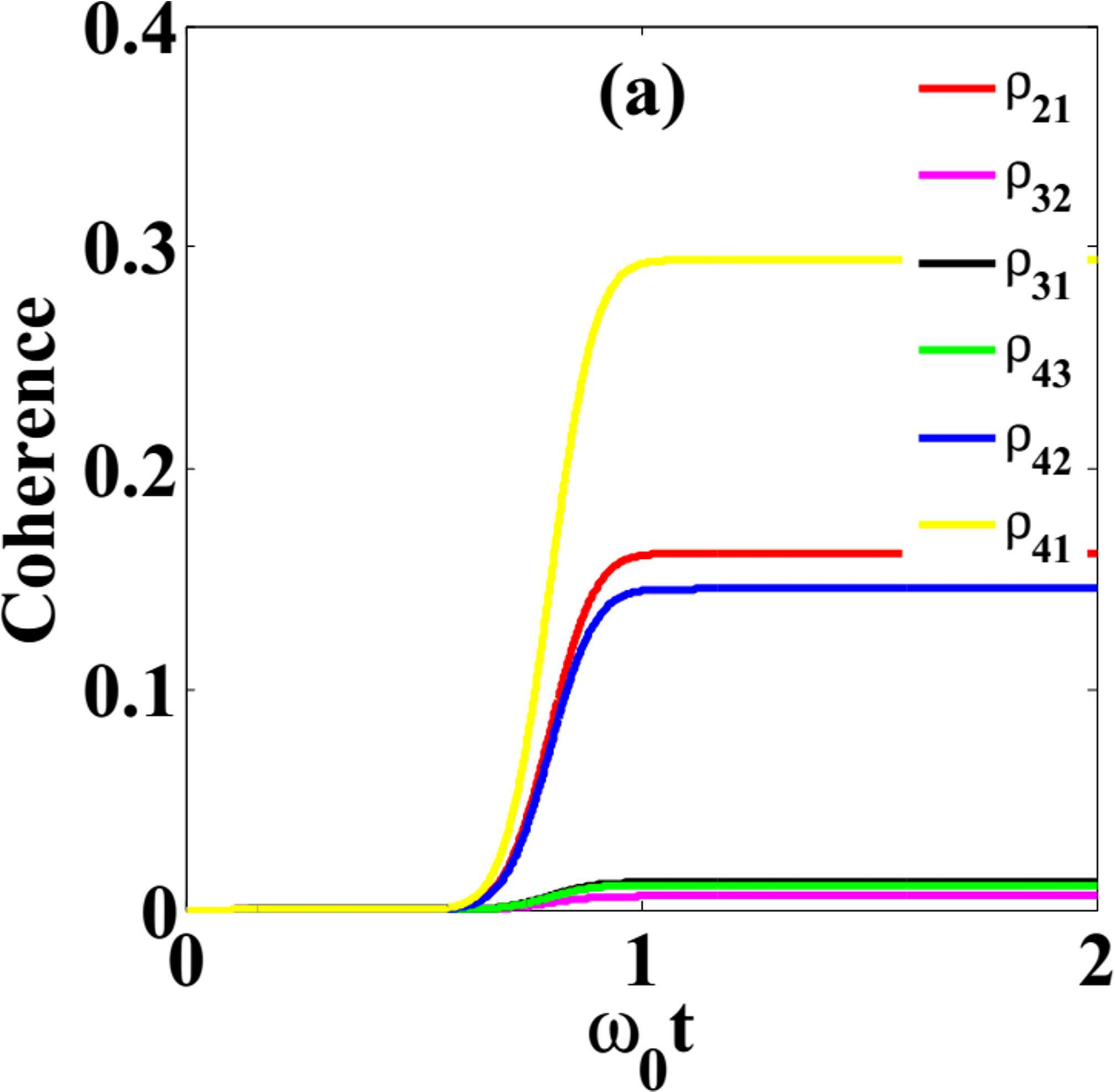}}
\quad
\subfloat{\label{Initial_Condition_Coherence1}
\includegraphics[width = 2.0 in]{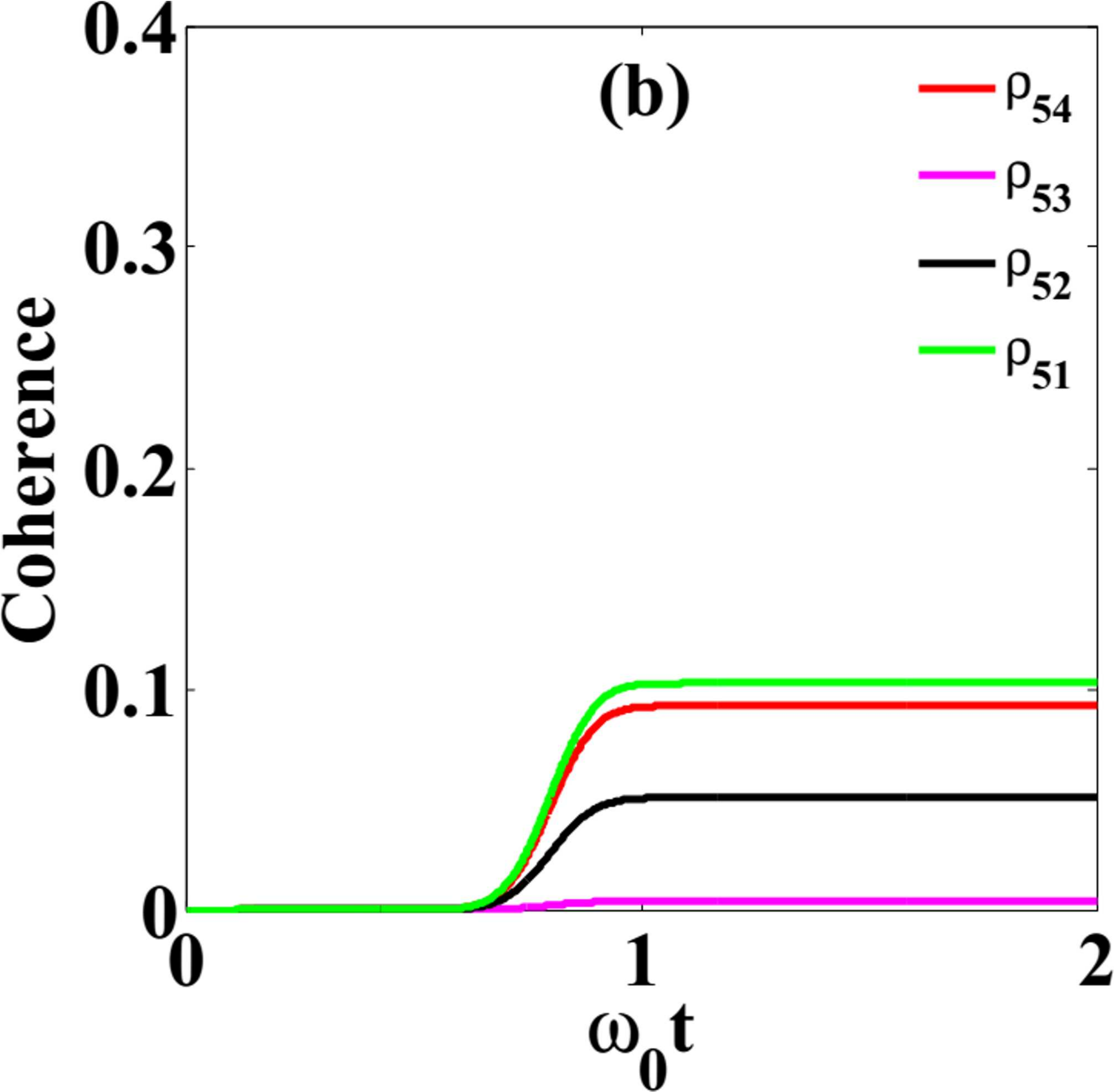}}
\quad
\subfloat{\label{Initial_Condition_Coherence2}
\includegraphics[width = 2.0 in]{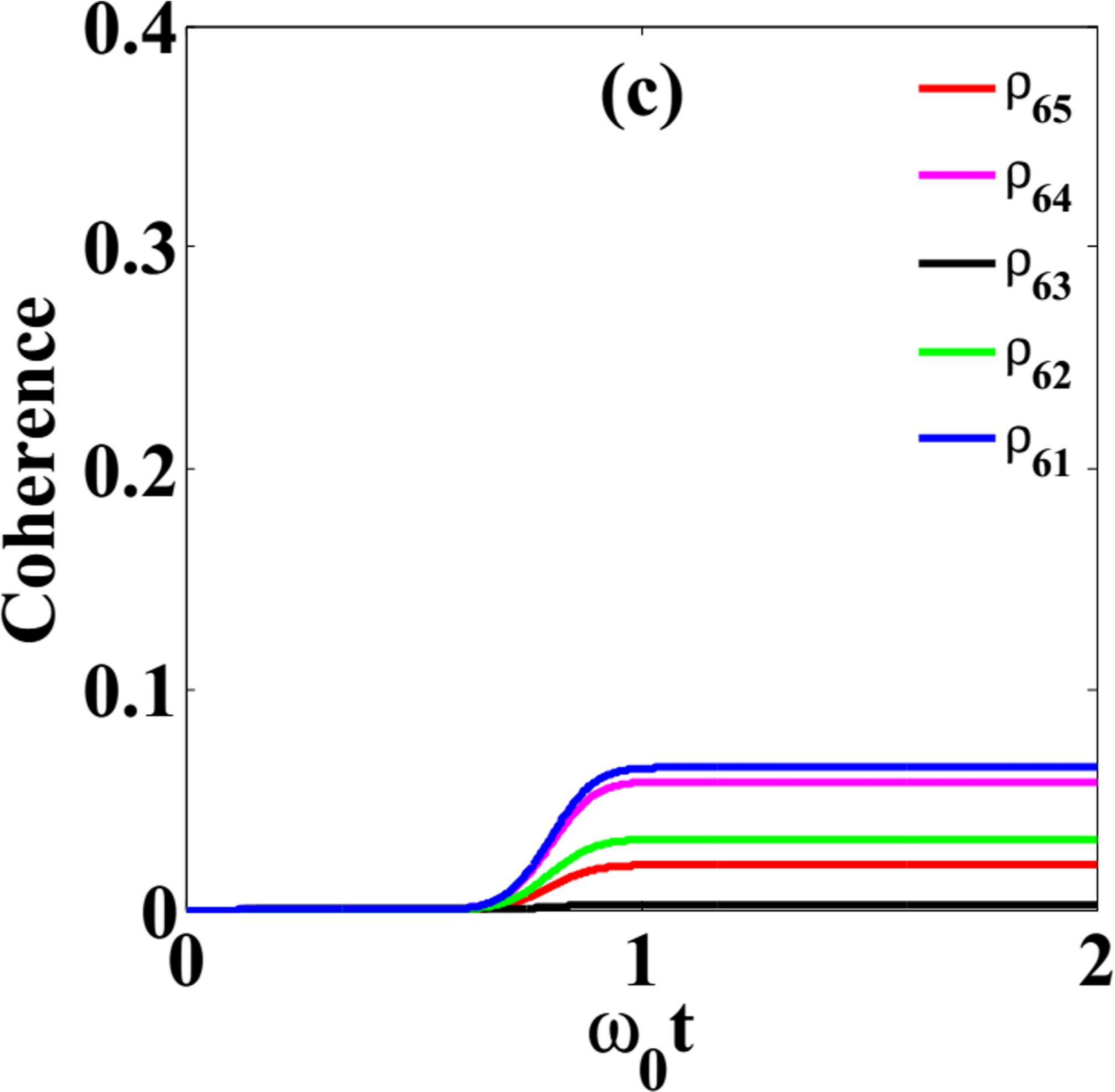}}
\quad
\subfloat{\label{Initial_Condition_Coherence3}
\includegraphics[width = 2.0 in]{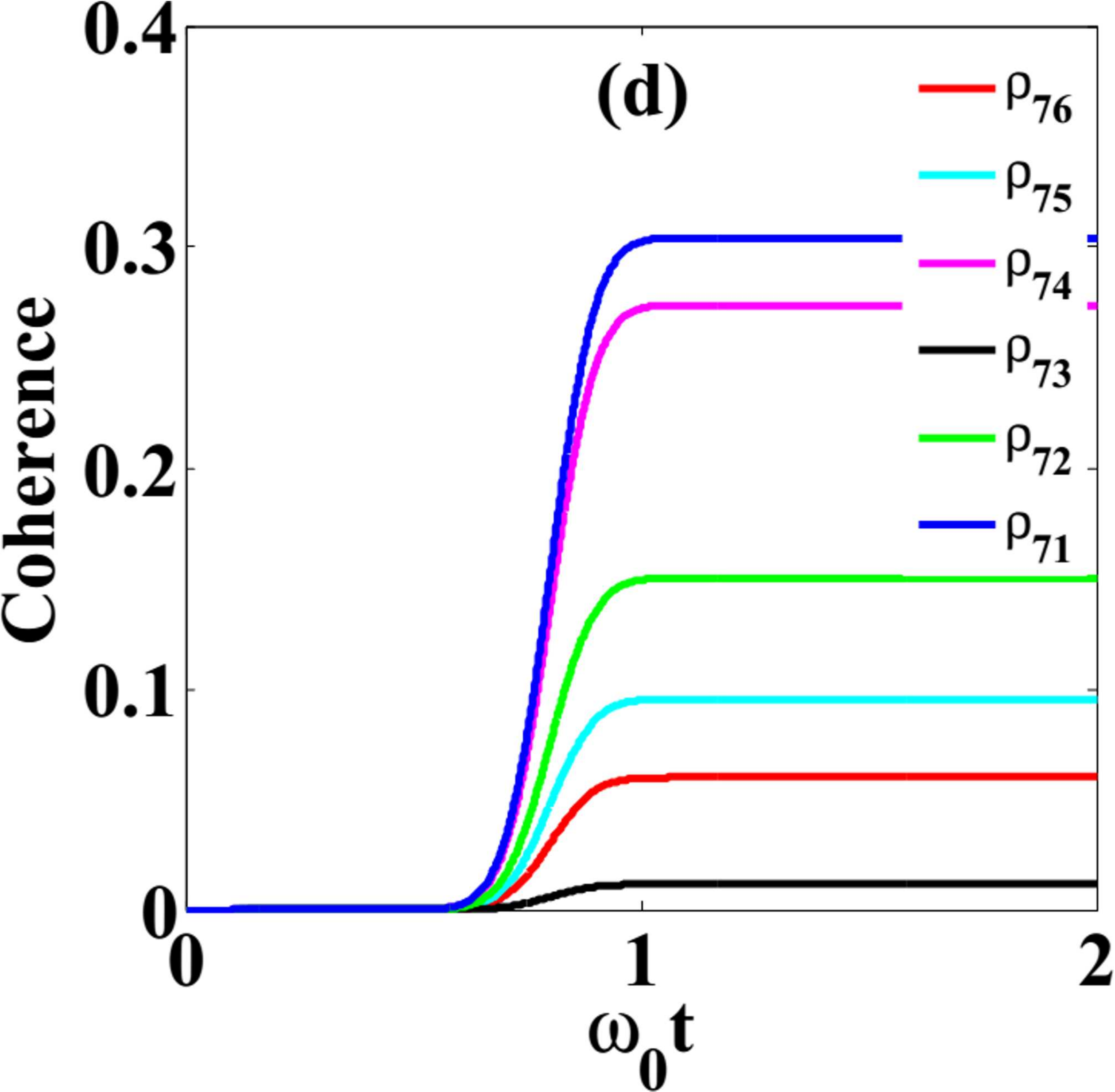}}
\quad
\subfloat{\label{Initial_Condition_Coherence4}
\includegraphics[width = 2.0 in]{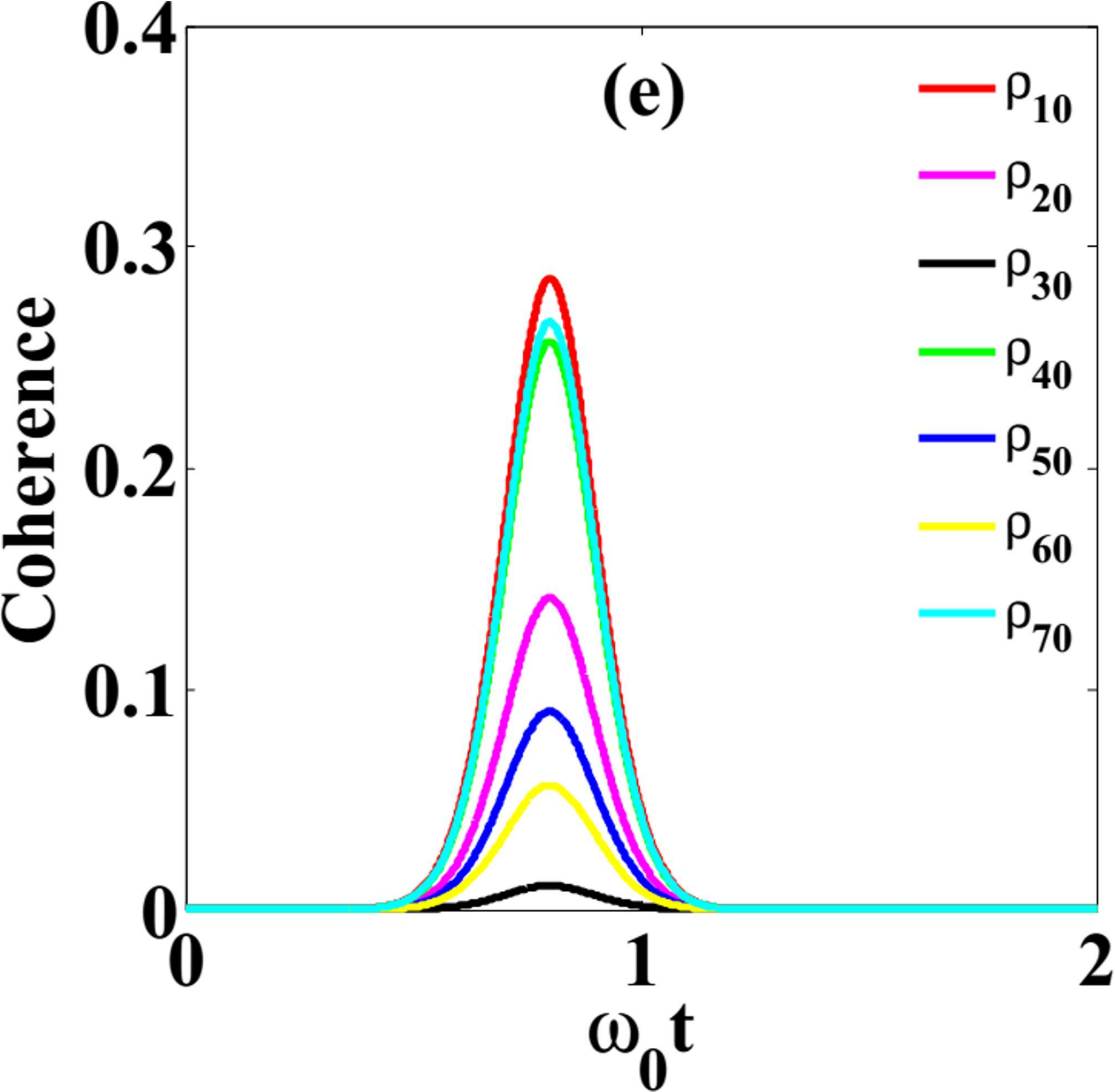}}
\caption{(Color online) Temporal dynamics of coherences between different BChla sites due to interaction with the coherent laser pulse.}
\label{Coherence}
\end{center}
\end{figure*}

Next, we study the dynamics of all the BChla sites in the presence of the pulse. In Fig. \ref{Pulse_Population}, we show that for suitable choices of the pulse parameters, one can pump all the population from the ground state to the excited states. In addition, as displayed in Fig. \ref{Coherence}, the coherence between the excited states and the ground state vanishes, while, the excited states build up certain nonzero coherence among themselves. It is particularly important here to note that almost after $60 fs$ (when the laser pulse has already interacted with the BChlas), as illustrated in Fig. \ref{Initial_Condition_Coherence4}, the ground-to-excited state coherence (i.e. $\rho_{j0} ;\qquad j\neq0$) dephases as has been observed experimentally by Panitchayangkoon {\it et al.} \cite{Panitchayangkoon_long_2010}. Clearly, a pulsed excitation creates a coherent superposition of all the excited states of the BChla sites, which then evolves under the action of Coulomb tunnelling and the bath modes, once they become {\it effective}. This is contrary to the assumption by Ishizaki and Fleming \cite{Ishizaki}, that it is either the BChla 1 or the BChla 6, that absorbs the excitation first, which then propagates through a chain of BChla sites to reach the reaction center. We emphasise that the unequal distribution of excitation across different BChla sites, as illustrated in Fig. \ref{Pulse_Population}, is primarily due to their different relative orientation of the angle of transition dipole moments with the pulse polarisation. 

On the other hand, according to Forster theory, one obtains an {\it in-situ} incoherent distribution of excitation at initial time (i.e t = 0) between baseplate BChla site and the FMO complex \cite{schmidt_am_busch_eighth_2011} as follows: 
\begin{equation*}
\rho_{11} = 0.6444; \rho_{22} = 0.2; \rho_{33} = 0.0; \rho_{44} = 0.0222; 
\end{equation*}
\begin{equation}
\rho_{55} = 0.0578; \rho_{66} = 0.0378;\rho_{77} = 0.0378; \rho_{ij} = 0.0 ;\qquad i \neq j. 
\label{In_Situ_initial}
\end{equation}

\begin{figure}[!h]
\begin{center}
\includegraphics[width = 3.0 in]{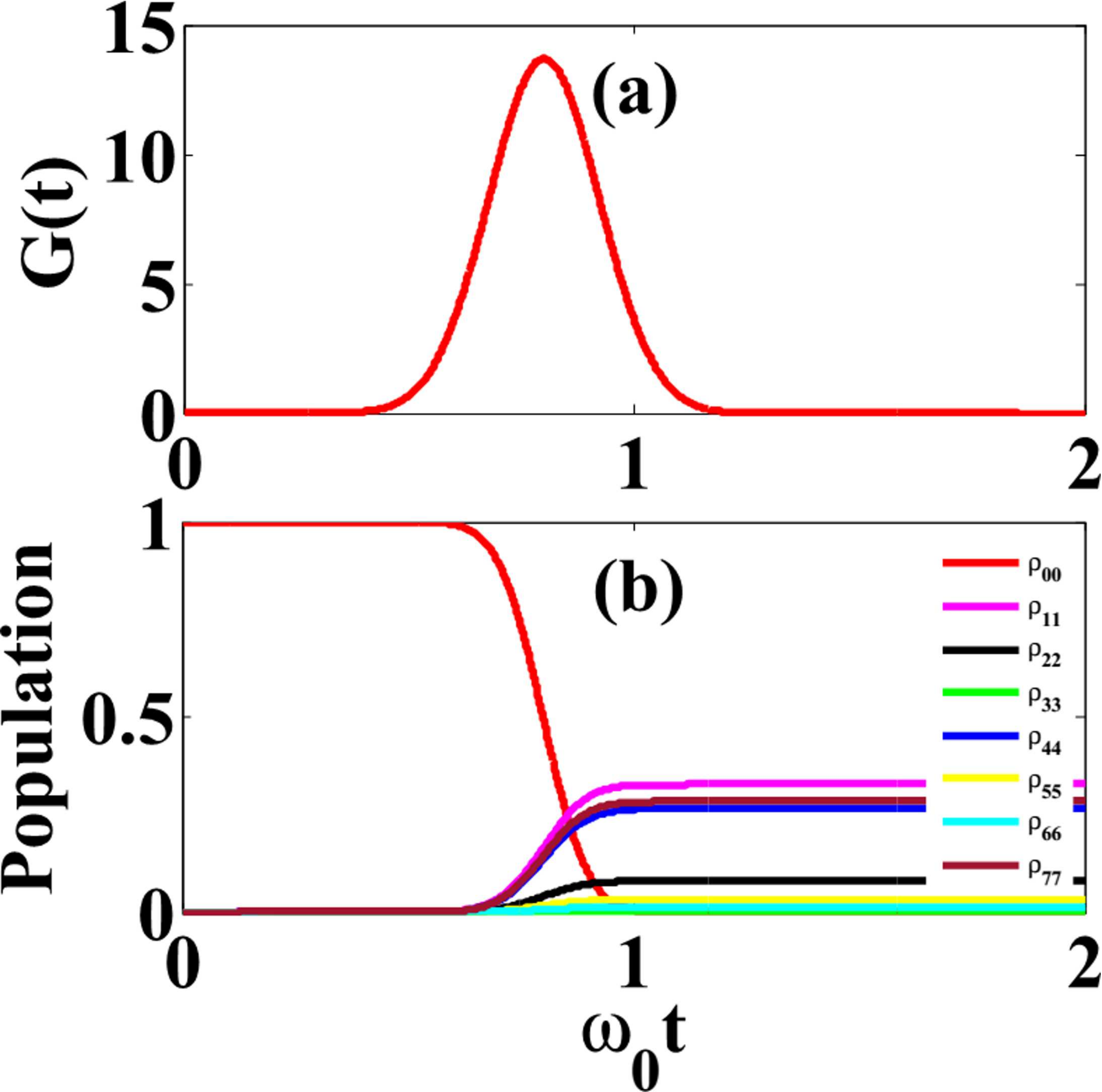}
\caption{(Color online) (a) Profile of Gaussian pulse of width $15$ fs. (b) Population dynamics of all the eight states of a monomer of the FMO complex, as an effect of a Gaussian pulse. We use normalized parameters $g = 13.6873$, $t_{0} = 0.8$ and $\Delta\tau^{2} = 0.03$.}
\label{Pulse_Population}
\end{center}
\end{figure}

\section{EET Dynamics}

\begin{figure}[!h]
\begin{center}
\includegraphics[width = 3 in]{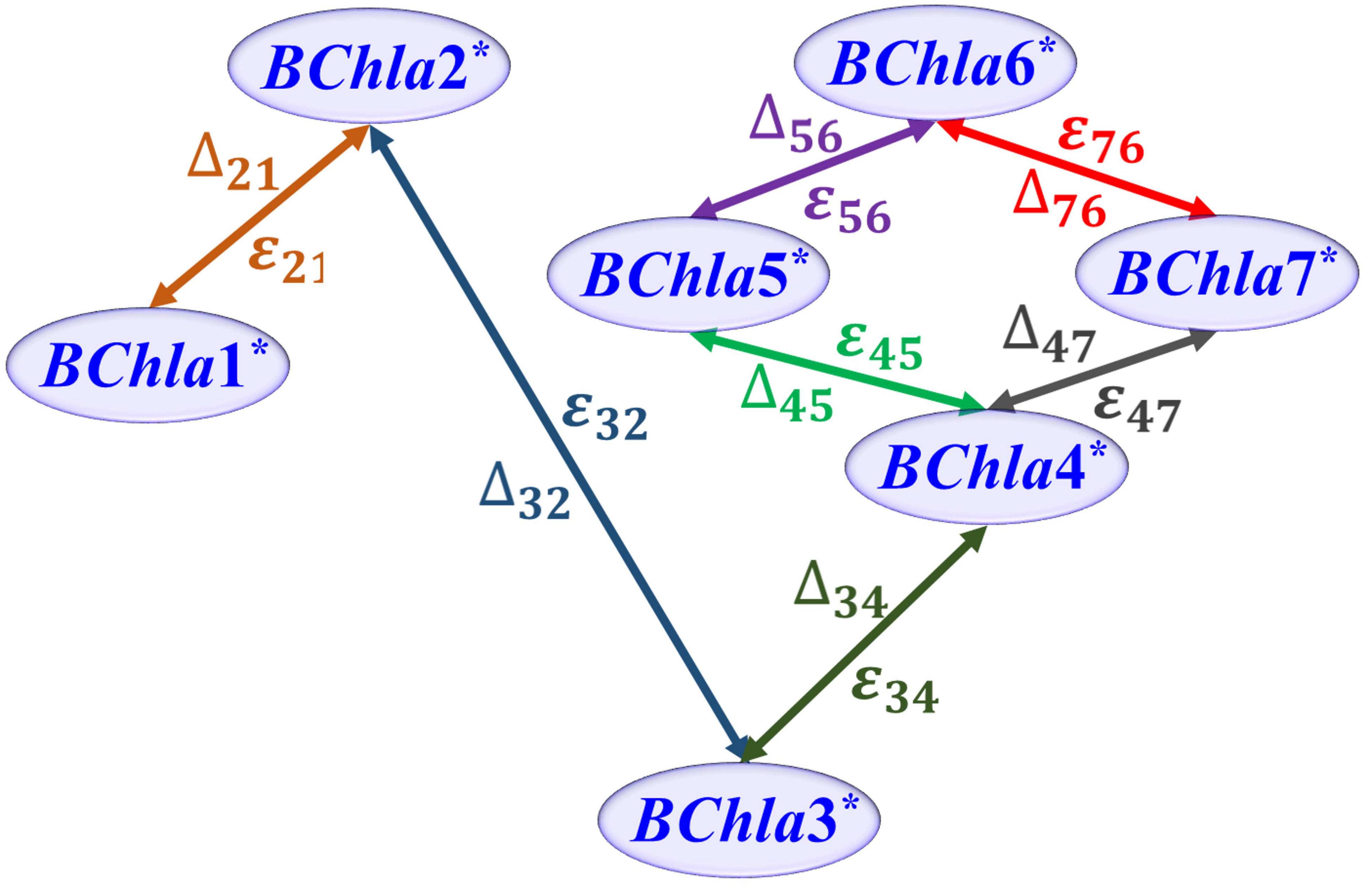}
\caption{(Color online) Schematic illustration of the active channels of EET in the monomer of FMO complex. The superscript '$\star$' indicates that the molecules are in excited state. Here $\epsilon_{ij} = \epsilon_{j} - \epsilon_{i}$ is the energy difference between $j$th and $i$th BChla sites, and $\Delta_{ij}$ represents the tunneling frequency between them.}
\label{Levels}
\end{center}
\end{figure}

In our numerical simulation of EET, we consider the relevant complex network of BChla sites in the monomer of FMO complex is shown in Fig.\ref{Levels}. We statrt with the total Hamiltonian, as \cite{Pachon,Gilmore}
\begin{equation}
H = H_{S} + H_{B} + H_{SB}\;,
\label{Hamiltonian}
\end{equation}
where   the system Hamiltonian $H_{S}$ is given by
\begin{equation}
H_{S} = \sum_{i,j}\left( \frac{\hbar}{2}\epsilon_{ij}\sigma_{z}^{ij} + \hbar\Delta_{ij}\sigma_{x}^{ij}\right)\;.
\label{System_Hamiltonian}
\end{equation}
Here $\sigma_{z}^{ij} = \ket{g_{j}e_{i}}\bra{g_{j}e_{i}} - \ket{e_{j}g_{i}}\bra{e_{j}g_{i}}$ and $\sigma_{x}^{ij} = \ket{e_{j}g_{i}}\bra{g_{j}e_{i}} + \ket{g_{j}e_{i}}\bra{e_{j}g_{i}}$ are the equivalent Pauli spin operators in two-site single excitation basis, $\epsilon_{ij} = \epsilon_{j} - \epsilon_{i}$ is the energy difference between $j$th and $i$th BChla sites, and $\Delta_{ij}$ represents the tunneling frequency between them as shown in Fig.\ref{Levels}.

The bath Hamiltonian can be written as
\begin{equation}
H_{B} = \sum\limits_{k_{ij}} \hbar \omega_{k_{ij}}b_{k_{ij}}^\dagger b_{k_{ij}}\;.
\label{Bath_Hamiltonian}
\end{equation}
Here, $b_{k_{ij}} = b_{k_{j}} - b_{k_{i}}$ and $b_{k_{ij}}^\dagger = b_{k_{j}}^\dagger - b_{k_{i}}^\dagger$ are the annihilation and the creation operators, respectively, for the $k_{ij}$th bath mode, where $b_{k_{i}}$ represents the $k$th bath mode, local to $i$th BChla site.

The system-bath interaction can be described by the spin-boson Hamiltonian as
\begin{equation}
H_{SB} = \frac{\hbar}{2}\sum_{i,j} \sum\limits_{k_{ij}} \sigma_{z}^{ij} g_{k_{ij}}(b_{k_{ij}} + b_{k_{ij}}^\dagger) \;,
\label{Interaction_Hamiltonian}
\end{equation}
where $g_{k_{ij}}$ is the electron-phonon coupling constant. 

To study the dynamics of EET, we use the following non-Markovian master equation for the system density matrix $\rho$ \cite{singh_coherence_2017,Carmichael}:
\begin{equation}
\begin{array}{lll}
\dot{\rho} &=& -\frac{i}{\hbar}\left[ H_{S},\rho\right]\\
&+&\frac{1}{4}\sum_{i,j}\left\{\left( \sigma_{z}^{ij}\rho\sigma_{z}^{ij} - \rho\sigma_{z}^{ij}\sigma_{z}^{ij}\right)D_{ij}(t)\right. \\
&+&\left( \sigma_{z}^{ij}\rho\sigma_{z}^{ij} - \sigma_{z}^{ij}\sigma_{z}^{ij}\rho\right)D_{ij}^{\ast}(t)\\
&+&\left( \sigma_{z}^{ij}\rho\sigma_{z}^{ij} - \sigma_{z}^{ij}\sigma_{z}^{ij}\rho\right)U_{ij}(t)\\
&+&\left.\left( \sigma_{z}^{ij}\rho\sigma_{z}^{ij} - \rho\sigma_{z}^{ij}\sigma_{z}^{ij}\right)U_{ij}^{\ast}(t) \right\} \;.
\end{array}
\label{non-Markovian}
\end{equation}
The time-dependent coefficients in Eq.(\ref{non-Markovian}) represent the system-bath correlations and are given by
\begin{eqnarray}
\label{correlation1}D_{ij}(t) &=& \int_{0}^{t}dt^{\prime}\int_{0}^{\infty} d\omega J_{ij}(\omega)\bar{n}(\omega,T)e^{-i\omega(t-t^{\prime})}\;,\\
\label{correlation2}U_{ij}(t) &=& \int_{0}^{t}dt^{\prime}\int_{0}^{\infty} d\omega J_{ij}(\omega)[\bar{n}(\omega,T) + 1]e^{-i\omega(t-t^{\prime})}\;,
\end{eqnarray}
where $\bar{n}(\omega,T)$ represents the average number of phonons and $J_{ij}(\omega)$ is the spectral density function considering the contributions from those vibronic modes corresponding to non-negligible values of Franck-Condon factors as originally proposed by Singh and Dasgupta \cite{singh_coherence_2017}. 

Here we also introduce the spectral function, as follows, that has been obtained by fitting with the experimental data (obtained using fluorescence line narrowing spectroscopy by Wendling {\it et al.\/}\cite{wendling_electronvibrational_2000})
\begin{equation}
J_{ij}(\omega) = K_{ij}\omega \left( \frac{\omega}{\omega_{c_{ij}}}\right)^{-1/2} e^{-\frac{\omega}{\omega_{c_{ij}}}} + \sum_l K_{l} e^{-\frac{(\omega - \omega_{l})^{2}}{2d^{2}}}\;.
\label{Spectral-Density}
\end{equation}
The above form of spectral density consists of the contributions of vibrational motion, arising from, for example,  the environmental phonons with the Huang-Rhys factor $K_{ij}$ (with $K_{ij} = g_{k_{ij}}^{2}$) and the vibronic modes with the Huang-Rhys factor $K_{l}$. Here $\omega_{c_{ij}}$ is the cutoff frequency and $\omega_l$ represents the frequencies of the active vibronic modes, with only the dominant Franck-Condon factors. In our analysis, we choose $\omega_{l}$ = 36 $cm^{-1}$, 70 $cm^{-1}$, 173 $cm^{-1}$, 185 $cm^{-1}$, and 195 $cm^{-1}$,  the values of $K_{l}$ equal to 40 times that of the corresponding Franck-Condon factors\cite{wendling_electronvibrational_2000}, and the width of the vibronic band as $d^2 = 18$. 

\begin{figure}[!h]
\begin{center}
\subfloat{\label{Comparison_6_77_exitons}
\includegraphics[width = 2.5 in]{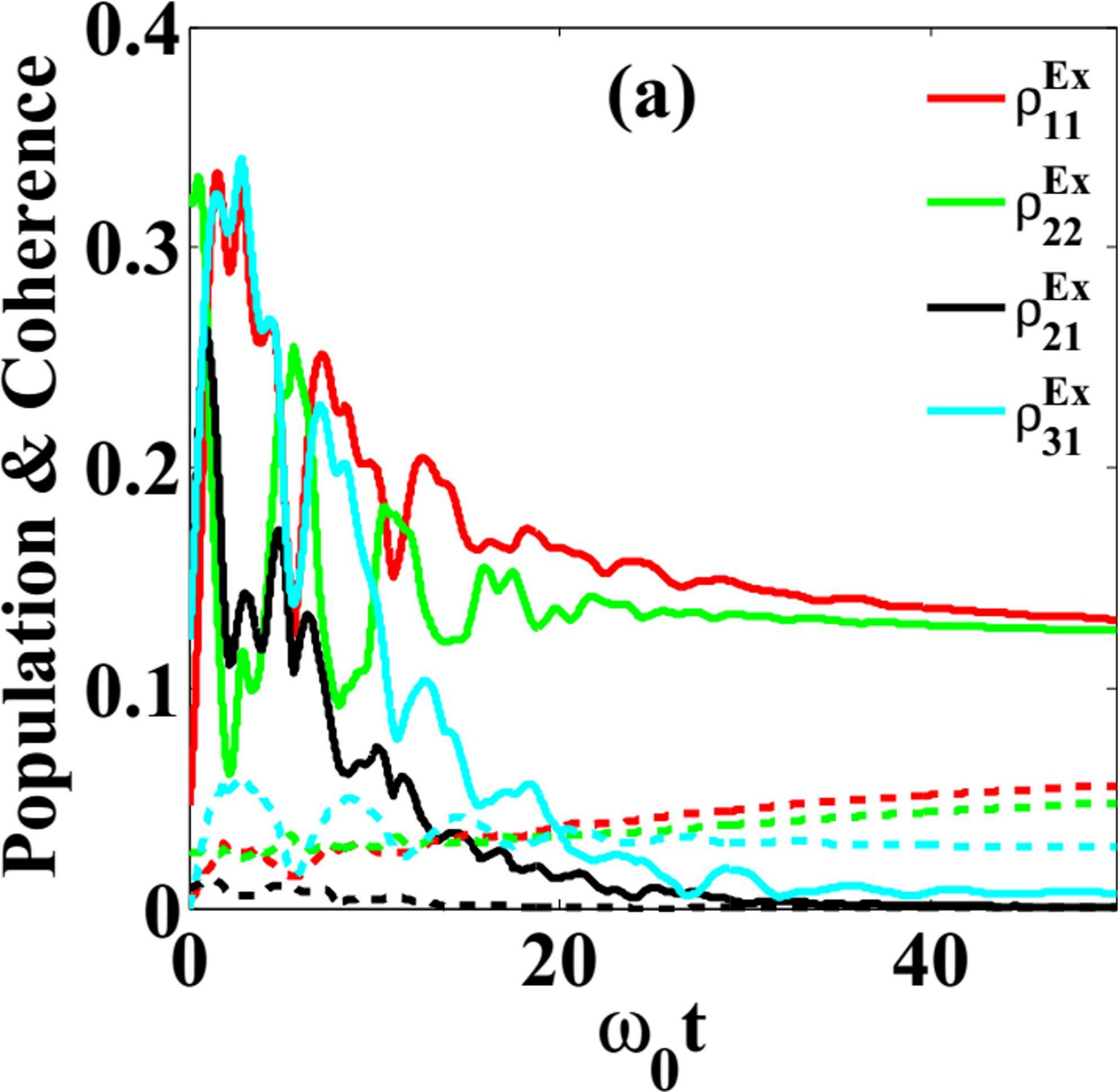}}
\quad
\subfloat{\label{Comparison_6_77_trace}
\includegraphics[width = 2.5 in]{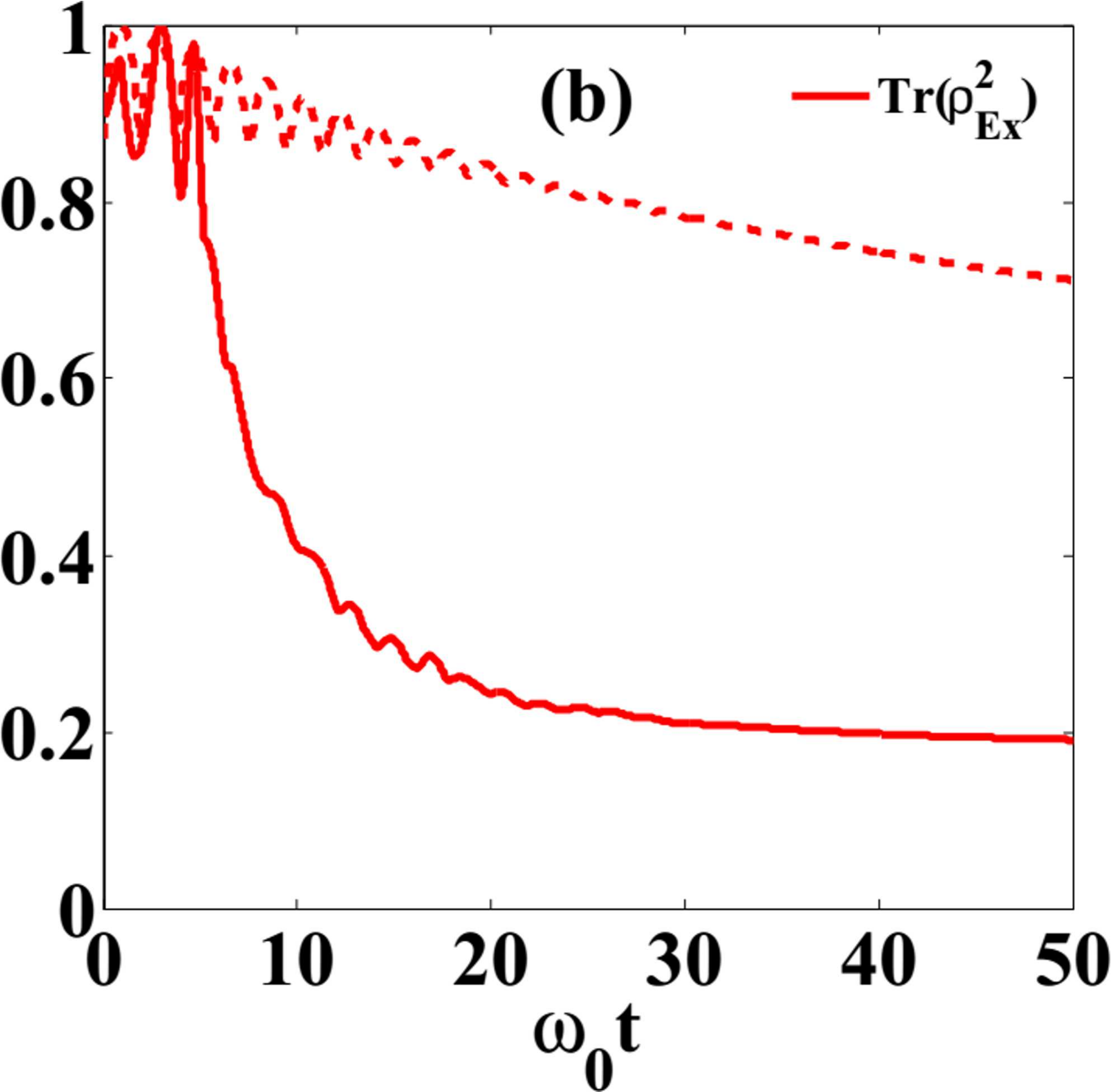}}
\caption{(Color online) Temporal dynamics of (a) the population and coherence and (b) the mixedness ${\rm Tr} (\rho^2)$ in the excitonic basis at a cryogenic temperature of 77 K. Thick lines show the dynamics for the initial condition obtained by the interaction of laser pulse with FMO complex, while dotted lines illustrate the dynamics for the initial condition of incoherent superposition determined using Forster theory.}
\label{Dynamics_Exciton}
\end{center}
\end{figure}

As in the 2-DES experiments, we choose the exciton basis. These excitons are delocalised on different BChla sites with time-independent probabilities. Following Cho {\it et al.} \cite{Cho}, we expand the density operator $\rho$ in the excitonic basis and numerically solve the Eq.(\ref{non-Markovian}) for following two different initial conditions:
\begin{enumerate}[(a)]
\item the state $|\psi(t)\rangle$, as obtained followed by the initial excitation by the laser pulse (see Fig. \ref{Pulse_Population} and \ref{Coherence}).
\item the mixed state given by Eq.(\ref{In_Situ_initial}) obtained according to the Forster theory\cite{schmidt_am_busch_eighth_2011}.
\end{enumerate} 
Here we plot only those four elements of density matrix for which the experimental information is available \cite{Engel_evidence_2007,Panitchayangkoon_long_2010,
panitchayangkoon_direct_2011}. For the Case (a), as illustrated by Fig.\ref{Comparison_6_77_exitons}, the frequency of oscillations of the relevant density matrix elements in the excitonic basis, namely $\rho_{31}^{Ex}$, $\rho_{21}^{Ex}$ and $\rho_{22}^{Ex}$, is almost the same as have been observed in the relevant experiments \cite{Panitchayangkoon_long_2010, panitchayangkoon_direct_2011}. More importantly, the time-scale till which the oscillatory nature persists, also matches with the experimental observations. In this case, the excitonic dynamical coherences $\rho_{21}^{Ex}$ and $\rho_{31}^{Ex}$  dephase in about 2 ps. Here it is interesting to note that even for the Case (b), excitonic population as well as coherence oscillates for about 700 fs as illustrated by Fig. \ref{Comparison_6_77_exitons}, but the time-scale of oscillations as well as the frequency of oscillations is considerably different from the experimental observations \cite{Panitchayangkoon_long_2010, panitchayangkoon_direct_2011}. It implies that consideration of more realistic initial condition is necessary to mimic the experimental results exactly.

Moreover, it is clear from the Fig. \ref{Comparison_6_77_exitons} that the dynamics of EET is quite fast for the case (a) as compared to the case (b). This is further confirmed by plotting the temporal profile of the mixedness ($\equiv {\rm Tr}(\rho^2)$) of the density matrix $\rho$   (Fig. \ref{Comparison_6_77_trace}). For the coherently prepared initial state [Case (a)]  the system achieves the steady state at around 2.5 ps  while for the initial incoherent distribution of excitation, the system is far away from the steady state even at 3 ps. To explore the reason behind such temporal behavior, we compare the dynamics of these two different initial conditions using Markovian as well as non-Markovian master equation.  Note that, in Markovian master equation the time-dependent correlation function of Eq.(\ref{correlation1}) and (\ref{correlation2}) become time-independent de-phasing rates $\gamma$ i.e $D(t) \rightarrow \frac{\gamma}{2}\bar{n}$ and $U(t) \rightarrow \frac{\gamma}{2}$. From our dynamical study we observe that the system attains steady state almost at the same time for both the initial conditions, in both the Markovian (Fig.\ref{Trace_Markovian_77_comparison}) as well as non-Markovian evolution (Fig. \ref{Comparison_6_77_trace}). It indicates that the speed up of coherently initiated dynamics is not due to non-Markovianity. This implies that the exchange of information between system and environment (as it happens in the non-Markovian dynamics), which preserves the coherence for long time, has no effect on the additional speed up of EET. Rather as the initial coherent superposition state relaxes more quickly than the mixed incoherent state, it is the initial coherence that plays the most crucial role in speeding up the EET dynamics.

\begin{figure}[!h]
\begin{center}
\subfloat{\label{Trace_Markovian_77_comparison}
\includegraphics[width = 3 in]{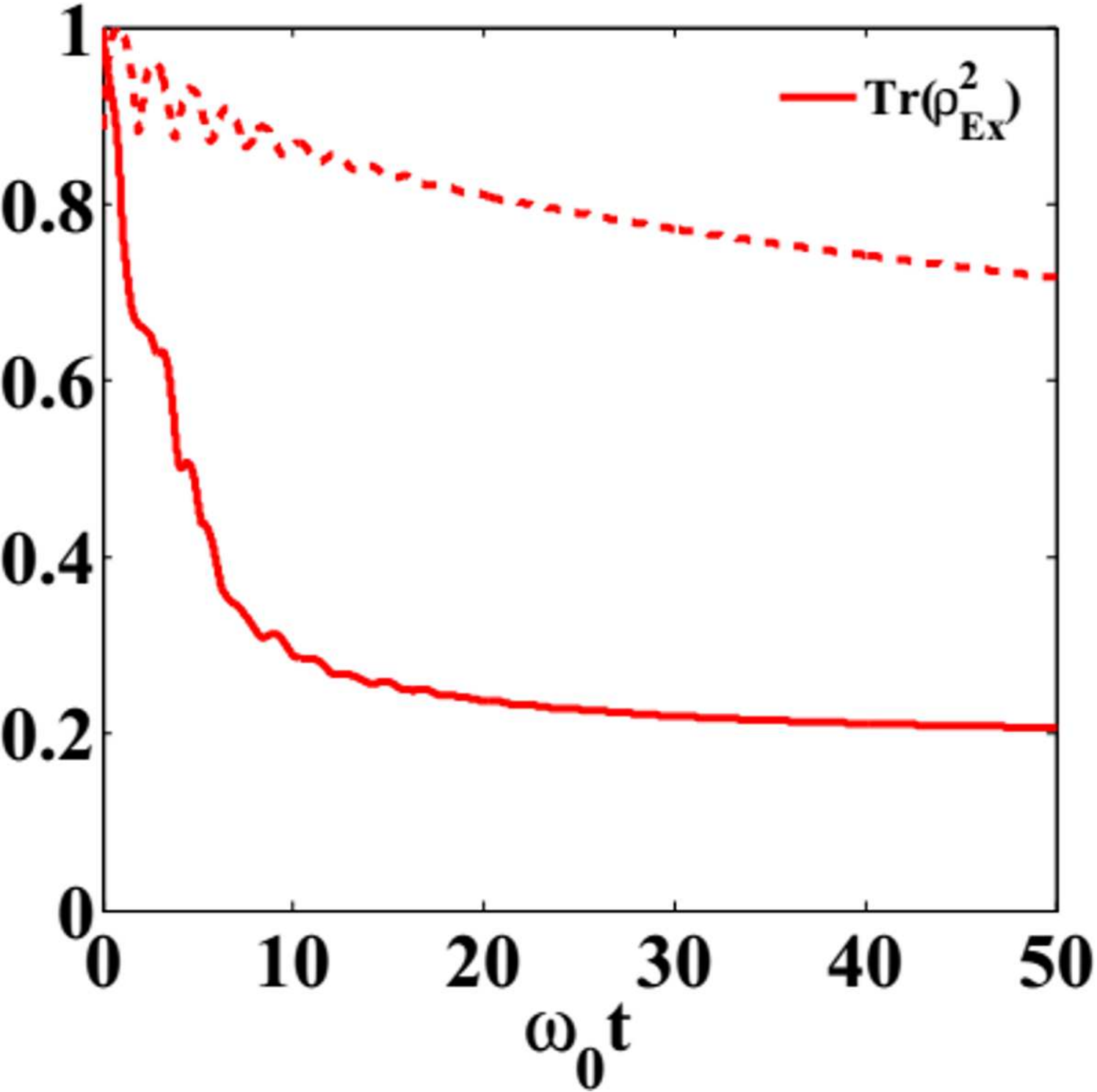}}
\caption{(Color online)Dynamics of trace of square of density matrix at cryogenic temperature 77 K with Markovian mater equation. Here thick line represents the coherently initiated dynamics while dotted line represents the incoherently initiated dynamics. We choose de-phasing rate $\gamma = 26 cm^{-1}$.}
\label{Trace_Markovian_77_comparison}
\end{center}
\end{figure}

To further elucidate the role of initial coherence we compared the dynamics of EET using two different initial coherent distributions; in first case we assumed the polarization of laser pulse parallel to transition dipole moment of BChla 1 and in the second case to that of BChla 6. From Fig.\ref{Comparison_1_6_77_trace} it is evident that in the latter case, the speed up of EET is more as compared to that in the former case. It implies that different initial coherent distributions leads to varied speed-up of EET in FMO complex. Similar observations of coherent control were investigated in detail for several molecular processes, e.g. photo-dissociation, bio-molecular reactions, by Shapiro, Brumer and their co-workers \cite{Shapiro}.

\begin{figure}[!h]
\begin{center}
\subfloat{\label{Comparison_1_6_77_trace}
\includegraphics[width = 3 in]{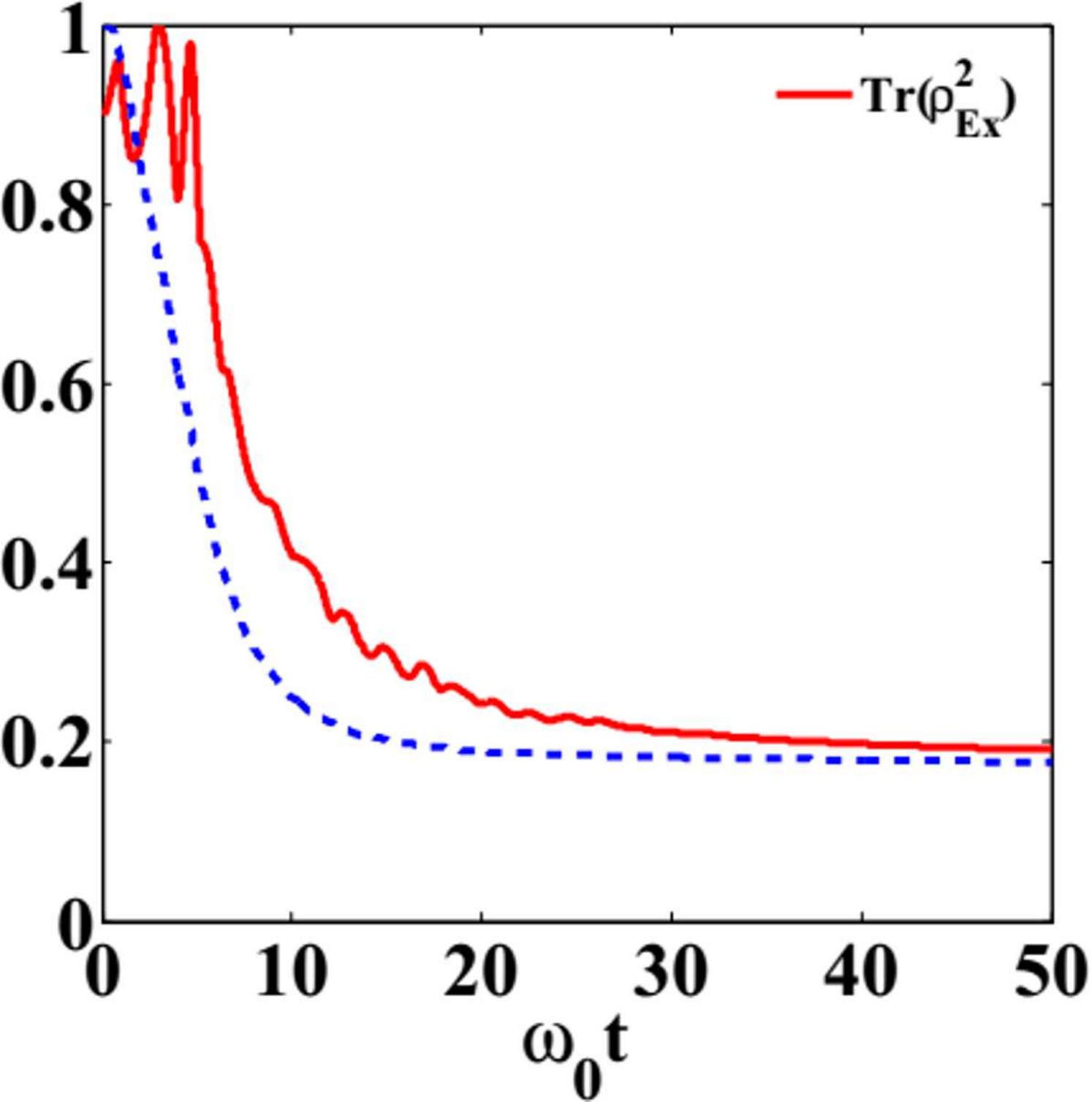}}
\caption{(Color online)Time evolution of trace of square of density matrix at cryogenic temperature 77 K. Solid line represents the initial condition when the polarization of laser pulse is parallel to transition dipole moment of BChla 1. Dashed line illustrate the initial condition when the polarization of laser pulse is considered to be parallel to transition dipole moment of BChla 6.}
\label{Comparison_1_6_77_trace}
\end{center}
\end{figure}

To get the further detail of role of initial coherence, we study the dynamics of EET by modifying the off-diagonal elements of the density matrix as discussed below. The evolution of the off-diagonal elements of the density matrix is governed by the equations of the form:
\begin{equation}
\dot{\rho_{ij}} = A\rho_{ii} + B(t)\rho_{ij}\;.
\end{equation}
Here the ad-hoc removal of the $B(t)$ term would refer to a case, when there is no initial coherence. Comparing the dynamics in the presence of $B(t)$ and in the absence of $B(t)$ (see Fig.\ref{Comparison_1_77_trace_Coherent}), we find that the system attains steady state much earlier in the presence of $B(t)$ than in the absence of $B(t)$.
\begin{figure}[!h]
\begin{center}
\subfloat{\label{Comparison_1_77_trace_Coherent}
\includegraphics[width = 3 in]{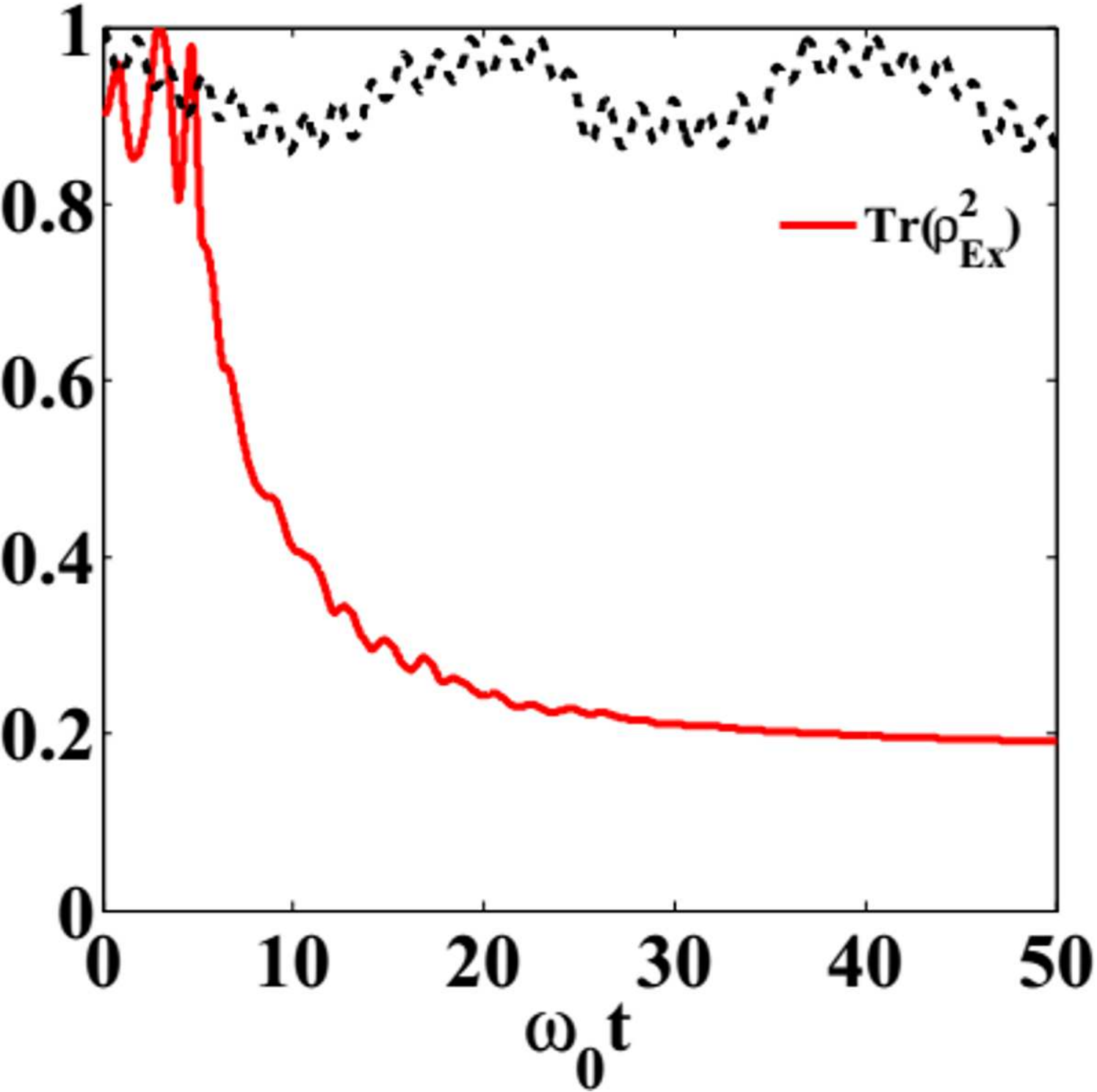}}
\caption{(Color online)Temporal dynamics of trace of square of density matrix at cryogenic temperature 77 K. Solid line represents the presence of $B(t)$, while dashed line represents the absence of $B(t)$.}
\label{Comparison_1_77_trace_Coherent}
\end{center}
\end{figure}

\section{Conclusions}

In conclusion, we have shown how the initial preparation of the collective state of the monomer can influence the EET. Precisely speaking, we have adopted an equivalent eight-state model of the singly excited BChla sites in a monomer and considered their relative orientation and energy differences. The dynamics of the EET is studied  using  master equation approach in the non-Markovian limit. It is found that the initial coherence of the state of the monomer takes a crucial role in the speed of EET.  This study suggests that the 2-DES experiments, in which one excites the monomer with a coherent pulse, does not properly mimic a realistic EET, in which the initial excitation is incoherent in nature. This also puts the two-pathway model of EET in FMO complex in question, in which it is assumed that either the BChla 1 or the BChla 6 are initially excited. 

\begin{acknowledgements}

This work was supported by Department of Science and Technology (DST), Govt. of India, under the grant number SR/S2/LOP-0021/2012. One of us (D.S.) gratefully thanks Paul Brumer, Jianshu Cao, Birgitta Whaley, Theodore Goodson \RNum{3} and Tomas Mancal for insightful discussions at QuEBS 2017, Jerusalem.

\end{acknowledgements}


\nocite{*}
\bibliography{Paper_Intial_Condition_Seven}     

\end{document}